\documentclass[%
 reprint,
 superscriptaddress,
 amsmath,amssymb,
 aps,
 floatfix
]{revtex4-2}

\usepackage{chemformula} 
\usepackage{xfrac}  
\usepackage{float} 
\usepackage{physics}

\usepackage{graphicx}
\usepackage{dcolumn}
\usepackage{bm}
\usepackage{xcolor}
\usepackage{soul}

\begin{document}


\title{Decoherence of Nuclear Spins in the Proximity of Nitrogen Vacancy Centers in Diamond}

\author{Mykyta Onizhuk}
    \affiliation{Department of Chemistry, University of Chicago, Chicago, IL 60637, USA}
\author{Giulia Galli}%
   \email{gagalli@uchicago.edu.}
   \affiliation{Department of Chemistry, University of Chicago, Chicago, IL 60637, USA}
   \affiliation{Pritzker School of Molecular Engineering, University of Chicago, Chicago, IL 60637, USA}
   \affiliation{Materials Science Division and Center for Molecular Engineering, Argonne National Laboratory, Lemont, IL 60439, USA}

\date{\today}

\begin{abstract}

Nuclear spins in the proximity of electronic spin defects in solids are promising platforms for quantum information processing due to their ability to preserve quantum states for a remarkably long time. Here we report a comprehensive study of the nuclear decoherence processes in the vicinity of the nitrogen-vacancy (NV) center in diamond. We simulate from first principles the change in the dynamics of nuclear spins as a function of distance and state of the NV center and validate our results with experimental data. Our simulations reveal nontrivial oscillations in the Hahn echo signal, pointing to a new sensing modality of dynamical-decoupling spectroscopy, and show how hybridization of the electronic states suppresses the coherence time of strongly coupled nuclear spins. The computational framework developed in our work is general and can be broadly applied to predict the dynamical properties of nuclear spins.

 \end{abstract}

\maketitle

\section{Introduction}
Nuclear spins in solids and molecules can preserve their quantum state for a remarkably long time, exceeding seconds \cite{Maurer2012, Bradley2022} and even hours \cite{Saeedi2013,Zhong2015}, compared to the typical millisecond timescale of electronic spin defects \cite{Bourassa2020, Wolfowicz2021}, due to their low magnetic moment. Hence nuclear spins are valuable resources for quantum information processing, including memory registers in quantum networks \cite{Pompili2021, Ruf2021, Hermans2022, Bradley2022}, nuclei-assisted quantum sensors \cite{PhysRevA.94.052330, Qiu2021} and components of fault-tolerant quantum processors \cite{Rong2015, PhysRevLett.117.130502}. In particular, in the presence of electron spin qubits in semiconductors and insulators, the hyperfine interactions between the electron and nuclear spins allow for electron-spin assisted initialization and read-out \cite{Ruskuc2022, Mdzik2022, Abobeih2022}, enabling full quantum control over the nuclear spin states.

Here, we consider the coherent lifetime of nuclear spins in the proximity of a state-of-the-art spin qubit platform, the negatively charged nitrogen vacancy in diamond \cite{DOHERTY20131, RevModPhys.92.015004} (NV). While the coherence properties of NV centers have been extensively investigated both experimentally \cite{Childress2006, PhysRevB.80.041201, PhysRevB.82.201201, PhysRevB.102.134210, PhysRevB.93.024305, PhysRevB.93.024305, Maze2012, BarGill2012} and theoretically \cite{PhysRevB.78.094303, PhysRevB.85.115303, PhysRevB.87.115122, PhysRevB.94.134107, Park2022}, our understanding of the nuclear spin qubit dynamics in the presence of an NV center is minimal. With scattered experimental data and qualitative analyses reported only for spin qubits in silicon \cite{PhysRevB.91.214303}, there is little consensus in the literature on the main physical mechanisms determining the decoherence of nuclear spins and how to control it. 

Acquiring a fundamental understanding of nuclear spin coherence in the proximity of electron spin qubits is crucial, e.g., to guide the design of nuclear spin environments for optimal performance of memory registers in quantum network applications \cite{Bradley2022, Wehner2018}. 


First principles simulations represent promising techniques to investigate decoherence of the spin qubits in solids. From predicting bath spin-induced relaxation \cite{Ivady2020, BulanceaLindvall2021}, identifying new host materials \cite{Kanai2022, Ye2019}, and sensing modalities \cite{Zhao2011, PhysRevB.101.155412, Ma2014, PhysRevB.105.075202} to engineering spin environments \cite{Onizhuk2021, PhysRevX.12.031028, Lee2022}, simulations have proved to be crucial in understanding spin-bath interactions in realistic systems. However to date, no attempt has been made to characterize nuclear spin coherence processes in the presence of a spin defect using accurate computational methods. Such a characterization is challenging as one needs to account for weak correlated fluctuations of numerous bath spins, where the dominant interaction arises from the electron-nuclear spin coupling.

In this Article, we use the cluster-correlation expansion (CCE) method in conjunction with density functional theory (DFT) to carry out \textit{ab initio} calculations of nuclear spin coherence dynamics in diamond in the vicinity of an NV center. Our computational results for nuclear spin Hahn-echo and Ramsey coherence times are in excellent agreement with experimental data, bridging the gap between theory and experiment. These calculations allow us to precisely assign the coherence times to the geometric positions of the nuclear spins relative to the spin defect, and identify the primary sources of nuclear spin decoherence in a wide range of conditions. Overall, our work provides a robust approach to predict nuclear spin coherence for a variety of systems.

\section{Results}

\subsection{Nuclear spins in a nuclear spin bath}

We begin by investigating the coherence properties of a \ch{^{13}C} nuclear spin in a \ch{^{13}C} nuclear spin bath of diamond with natural isotopic abundance in the absence of any electron spin (Fig. 1a). We adopt two theoretical frameworks - the CCE approach \cite{PhysRevB.78.085315}, which assumes that decoherence arises only from dephasing, and the generalized CCE approach (gCCE) \cite{Onizhuk2021c}, which accounts for both relaxation and dephasing of the central spin (see Methods). 

\begin{figure}
    \centering
    \includegraphics[scale=1]{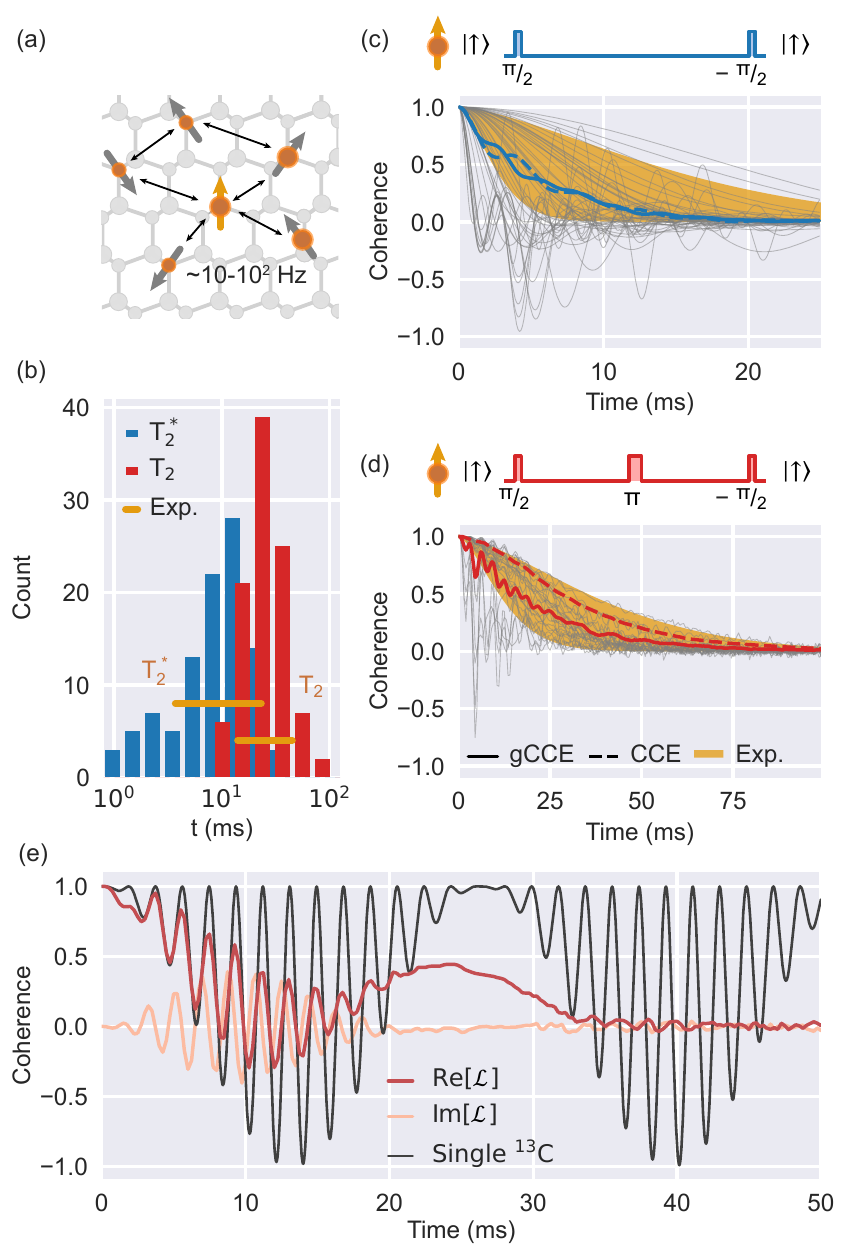}
    \caption{(a) Schematic representation of a nuclear spin in a nuclear spin bath. (b) Distribution of Ramsey ($T_2^*$) and Hahn-echo ($T_2$) coherence times for nuclear spins computed with the gCCE approach (see Methods). Horizontal bars show the range of experimental values\cite{10qubitreg}. (c (d)) Computed Ramsey (Hahn-echo) signals of the nuclear spins. The top diagram represents the sequence of pulses for each type of  experiment. Each grey trace was obtained for a single random configuration and computed at the gCCE level of theory. Colored lines show ensemble-averaged coherence curve computed with the CCE (dashed line) and gCCE (solid line) methods. The applied magnetic field is 50 mT. (e) Real (red) and imaginary (orange) part of the Hahn-echo coherence function for a random bath configuration which contains a bath spin coupled to a central spin with $\sigma=151$ Hz (Eq. \ref{eq:hmodel}) at a magnetic field of 0.05 mT. Analytical expression for $\langle\hat I_x\rangle$ of the system with only the central spin and a single bath spin shown in black (Eq. \ref{eq:ix}).}
    \label{fig:nonly}
\end{figure}

Figure \ref{fig:nonly}b shows the computed coherence time of the nuclear spins corresponding to Ramsey and Hahn-echo measurements. We find an excellent agreement between theory and experiment \cite{10qubitreg}. Ramsey calculations (Fig. \ref{fig:nonly}c) converge at the 2nd order of the CCE (CCE2) and Hahn-echo results (Fig. \ref{fig:nonly}d) converge at the 4th order (See SI). The difference between Ramsey signals computed at the first and second order is small, indicating that the single bath spin dynamics dominates the decoherence process, as expected \cite{PhysRevB.85.115303}. The order at which the Hahn-echo signal converges is significantly higher than that typically required to investigate the coherence time of electron spins (CCE2)\cite{Seo2016, PhysRevLett.129.117701, Jahn2022}, highlighting the need to account for higher-order correlations of the bath dynamics to accurately predict nuclear spin coherence times. 

We note that the distributions of the inhomogeneous $T_2^*$ and homogeneous spin dephasing times $T_2$ overlap in Fig. 1b, and the coherence enhancement from the refocusing pulse is on average small, a characteristic behavior of a broad noise spectrum \cite{Yang_2016}.

However, the behaviour of the coherence function turns out to be much more complex than one might expect from classical stochastic noise models, where inter-nuclear interactions are simply treated as an effective nuclear spin field \cite{PhysRevB.99.205423}. Only with a complete quantum-mechanical treatment can we uncover the complex oscillatory dynamics of the Hahn-echo signal (Fig \ref{fig:nonly}d), which has not been reported before for solid-state spin qubits. The oscillations arise from the direct spin-exchange interactions with single spins in the environment: if one neglects the spin exchange (CCE framework), the coherent oscillations are not present in the Hahn-echo signal (Fig. \ref{fig:nonly}d). Similar effects have been observed in the electron spin-echo modulation (ESEEM) of electron-radical pairs in organic molecules \cite{Salikhov1992, Kulik2001}. Contrary to the ESEEM arising from perpendicular hyperfine couplings \cite{Seo2016}, these oscillatory features do not disappear with increasing magnetic field.


One can use the spin-echo sequences to extract the spin-exchange couplings between nuclear spins. Here, we derive an analytical expression for the Hahn-echo coherence function $\mathcal{L}(t)=\frac{1}{\mathcal{L}(0)}\left(\langle \hat I_{x,0}(t) \rangle+i\langle \hat I_{y,0}(t) \rangle\right)$ of a central spin coupled to a single bath spin. We consider the following simplified Hamiltonian of the system:
\begin{equation}\label{eq:hmodel}
    \hat H = w_0 \hat{I}_{z,0} + w_1 \hat{I}_{z,1} + \frac{1}{2}\sigma(\hat{I}_{+,0}\hat{I}_{-,1}+h.c.)
\end{equation}
where $\hat{I}_{i,0}$ and $\hat{I}_{i,1}$ are spin operators for the central spin and the bath spin respectively, $w_0$, $w_1$ are Larmor frequencies, and $\sigma$ is the spin-exchange coupling. The presence of $\hat{I}_{z,0}\hat{I}_{z,1}$ type of couplings in the Hamiltonian leads to the same expression for the spin magnetization (see Eq. \ref{eq:ix} and \ref{eq:iy} below) and hence they were omitted in Eq. \ref{eq:hmodel}.
Assuming the initial state of the central spin is $\ket{+X}=\frac{1}{\sqrt{2}}(\ket{\uparrow}+\ket{\downarrow})$ and the $\pi$-pulse applies a rotation around the $x$-axis, we obtain the following expressions for the spin magnetization:

\begin{equation} \label{eq:ix}
\begin{split}
    \langle \hat I_{x,0} (t) \rangle =   \frac{1}{2} - \frac{\sigma^2}{4\Omega^2}&(
         1  - \cos{(\frac{\Omega t}{2})} 
           - \cos{(\frac{(w_0 + w_1)t}{2})} \\ 
          & + \cos{(\frac{(w_0 + w_1)t}{2})}\cos{\frac{\Omega t}{2})}),
\end{split}
\end{equation}
and
\begin{equation}\label{eq:iy}
    \langle \hat I_{y,0}(t) \rangle = \frac{\sigma^2}{2\Omega^2}
        \sin^2{(\frac{\Omega t}{4})}\sin{((w_0 + w_1)\frac{t}{2})},
\end{equation}
where $\Omega=\sqrt{(w_1-w_0)^2+\sigma^2}$. For a \ch{^{13}C} nuclear spin in  a \ch{^{13}C} spin bath, Larmor frequencies are equal, $w_0=w_1=-\gamma_n B_z$, and the spin-exchange coupling arises from the dipolar interactions $\sigma=-P_{zz}$ (see Methods).

Hence, one can observe a strong out-of-phase signal $ \langle \hat I_y(t) \rangle $ (Eq. (\ref{eq:iy}) and Fig. \ref{fig:nonly}e), which should be easily detectable in the experiment, providing a new way to directly measure spin-exchange coupling between spins in solids. In contrast to existing methods that probe the nuclear spin pair dynamics with the sensor spin \cite{Abobeih2018, PhysRevX.12.011048}, the echo oscillations characterized here require selective $\pi$-pulses and readout on one of the spins, but do not necessitate an auxiliary probing qubit. 

In addition, our calculations show that the role of longitudinal relaxation in determining the decoherence processes is highly dependent on the given nuclear spin configuration. Relaxation is negligible for some nuclear spins; for others, it completely determines the decoherence rate (See SI for examples). On average, the spin exchange with the environment accounts for about 40\% of the decoherence rate for the ensemble of nuclear spins (gCCE ensemble $T_2$ 22.4(2) ms vs. CCE 36.66(9) ms) and 30\% for single ones. Thus, for each specific electronic spin defect present in a solid, one should perform a detailed search within all experimentally available nuclear spins to identify the ones best suited for quantum memories.


\begin{figure}
    \centering
    \includegraphics[scale=1]{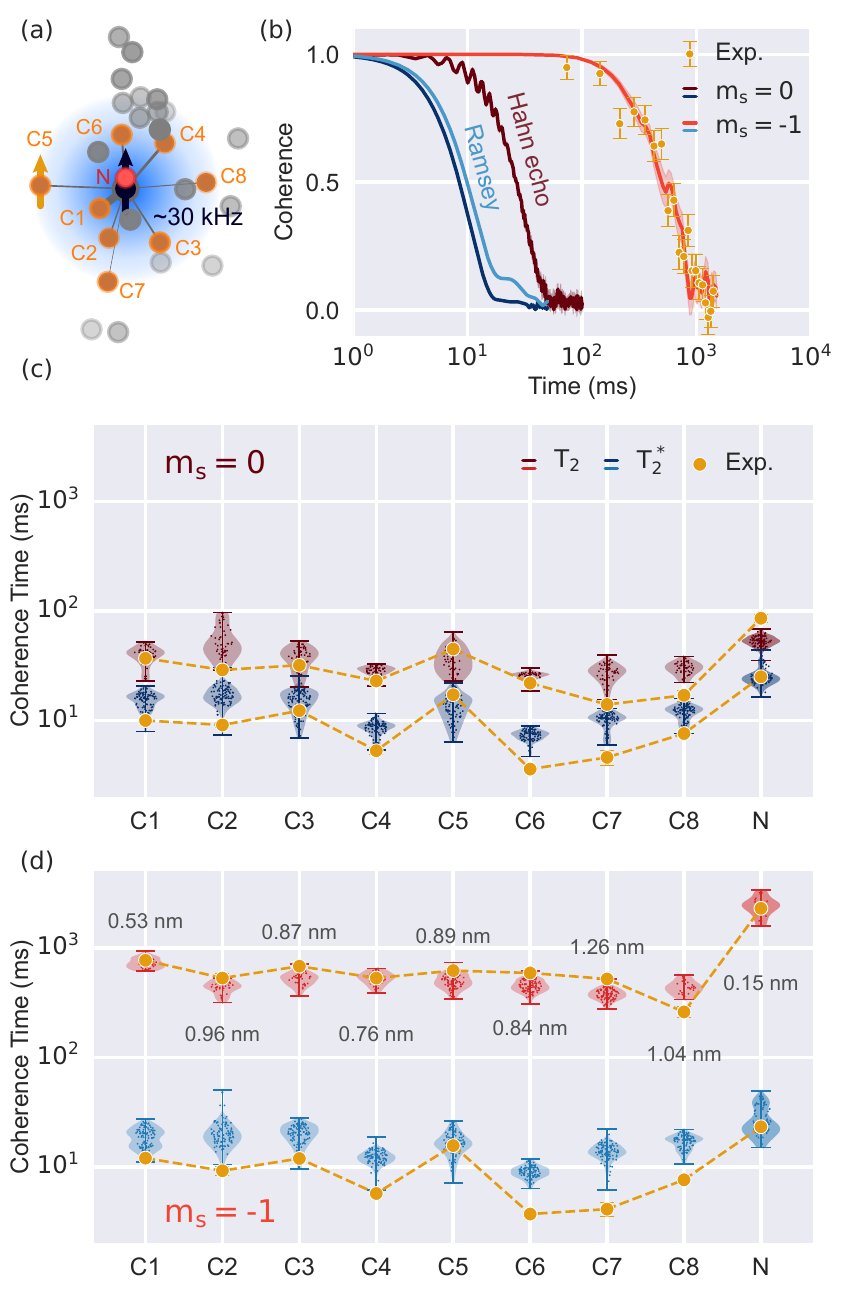}
    \caption{ (a) Graphical representation of the experimentally determined positions of 27 \ch{^{13}C} nuclear spins in proximity of an NV center in diamond from ref. \cite{Abobeih2019}. Orange circles show nuclear spins with measured coherence times. (b) Coherence signals for the nuclear spin C5. Solid lines are theoretical predictions; yellow points are experimental data. (c (d)) $T_2$ and $T_2^*$ of the nine nuclear spin registers measured by Bradley et al. \cite{10qubitreg} and represented by yellow lines when the NV is in the $m_s=0$ ($m_s=-1$) state. Distributions correspond to computed coherence times in 50 random nuclear spin configurations around the 27 nuclear spins, identified in the experiment (see text). The Hahn echo is computed at the 4th (5th) order of the cluster expansion for the \ch{^{13}C} (\ch{^{14}N}) nuclear spins.}
    \label{fig:validation}
\end{figure}

\begin{figure*}
    \centering
    \includegraphics[scale=1]{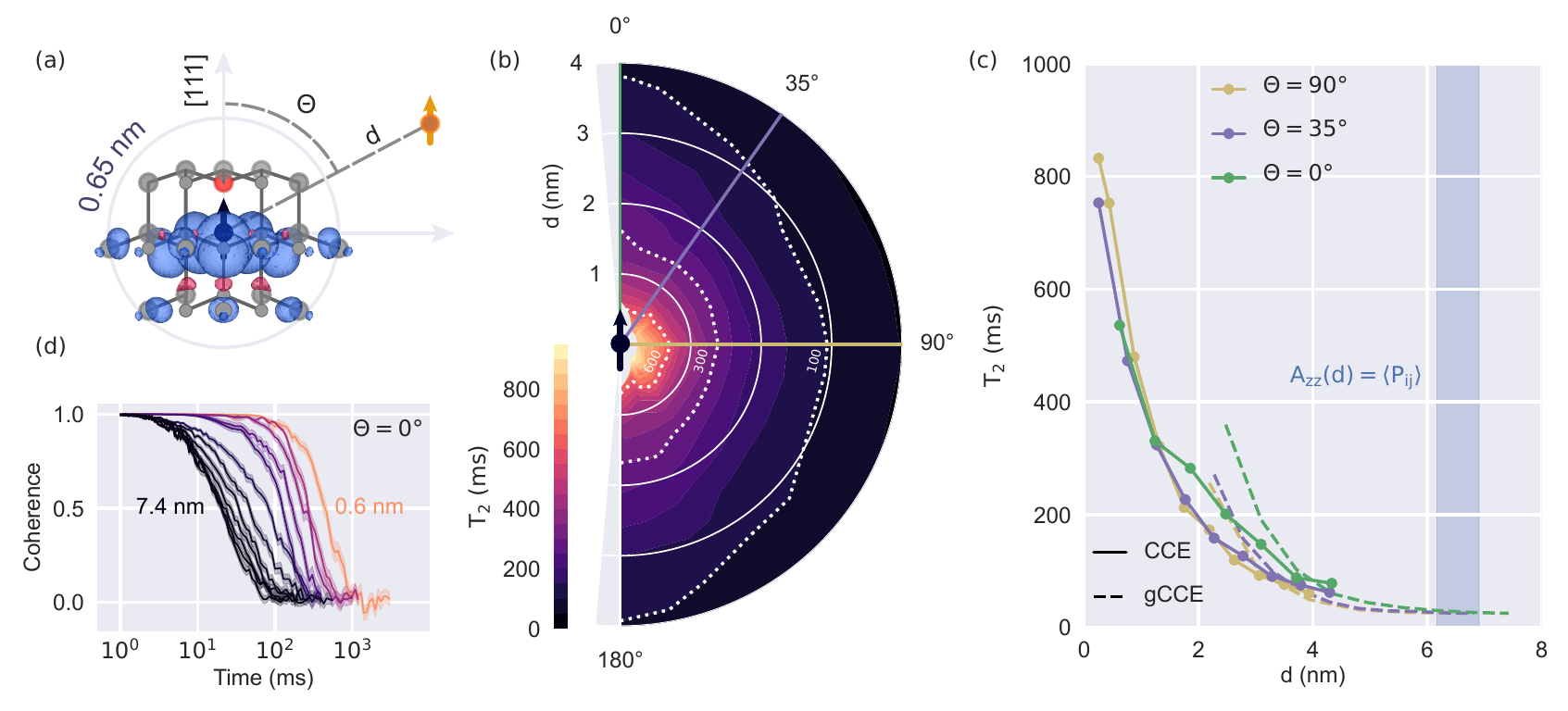}
    \caption{(a) Spin density distribution around the NV center in diamond computed using DFT and the PBE functional. $d$ is the distance from the NV center, the polar angle $\Theta$ is the angle between the NV axis and the position of the nuclei.
    (b) Computed ensemble-averaged $T_2$ as a function of the distance $d$ from the NV center and the polar angle $\Theta$. The coherence time of the nuclear spins at distances $\leq 0.5$ nm computed at the magnetic field of 1 T to decouple electron and nuclear spin, for all others distances $B=50$ mT. Dashed white lines show distances at which the $T_2$ is (from left to right) 600, 300, and 100 ms.
    (c) Computed $T_2$ of the nuclear spin at three polar angles $\Theta=0^\circ, 35^\circ, 90^\circ$. The generalized cluster expansion (gCCE) simulations (dashed line) were carried out using a smaller number of clusters than CCE, converged for free nuclear spin (see SI).
    The value of d where the mean hyperfine coupling is equal to the mean internuclear coupling $A_{zz}(d)=\langle{P_{ij}}\rangle$ was computed as $\left| \frac{\gamma_e}{\gamma_n} \overline{r}_{ij}^3 \right|^{\sfrac{1}{3}}$ where $\overline{r}_{ij}$ is the mean internuclear distance (0.45-0.5 nm).
    (d) Computed ensemble-averaged Hahn echo as a function of the distance from the NV center for nuclear spins aligned along the $[111]$ direction.}
    \label{fig:t2map}
\end{figure*}


\subsection{Experimental validation of the computational protocol in the presence of an electron spin}
Now we turn our attention to the properties of nuclear spins in the presence of an NV center in diamond.
With a total spin of 1, the NV center can be initialized in three eigenstates, which differ by the projection of the magnetic moment along the [111] axis of diamond ($m_s = -1,0,1$).
By preparing the NV in the $m_s = 0$ state, one can, up to first order, eliminate the electron spin coupling to the spin bath and recover the same coherence time that nuclear spins exhibit in a pure nuclear spin bath (i.e, free nuclear spins). In the $m_s = -1,1$ states, the NV center induces a strong hyperfine field on the nuclear spins, which dominates the inter-nuclear interactions. The hyperfine field gradient greatly suppresses the polarization transfer between different nuclei, leading to a significant change in the nuclear spin dynamics - an effect known as frozen core \cite{Neumann2013, PhysRevB.91.214303}.

We validate the predictions of our calculations by comparing our results with the experimental measurements of coherence times reported by T. H. Taminiau and coworkers \cite{10qubitreg, Abobeih2019, Abobeih2022}. The data for NV in $m_s = 0$ presents a new and previously unpublished data set obtained on the same NV center as used in these studies (C. E. Bradley and T. H. Taminiau, personal communication, 2023). To apply $\pi$-pulses to the separate nuclear spins in the experiment, one has to include a short period of time during which the electron spin is in $m_s=-1$ state (see Methods), which might lead to small discrepancies between the theoretical predictions and the experimental data.

We prepare a set of random configurations of nuclear spins placed around a cluster of 27 nuclear spins with experimentally identified positions \cite{Abobeih2019} and compute the coherence of nine selected nuclear spins (C1-C8 and N in Fig. \ref{fig:validation}a). Using \textit{ab initio} computed hyperfine parameters for randomly placed nuclear spins (see Methods), we obtain both Hahn-echo and Ramsey coherence times of all nine nuclear registers. 

Our calculations show that the presence of the electronic defect center greatly affects the nuclear spin qubit dynamics under the dynamical decoupling protocol.
For example, in Fig. \ref{fig:validation}b, we show that the Hahn-echo coherence time of the C5 nuclear register is enhanced by a factor of 18 when the electron spin is in the $m_s = -1$ state. We also find a clear correlation between distance from the NV and the $T_2$ of the nuclear spins. Maximum $T_2$ values are achieved for the \ch{^{14}N} nuclear spin, which is located in the center of the frozen core and has a lower gyromagnetic ratio than that of \ch{^{13}C}.

The electron-nuclear spin interactions dominate the dynamics of the nuclear spin bath; thus, an accurate description of the nuclear spin's decoherence processes requires accounting for numerous weak correlated fluctuations of the bath spins. For the CCE calculations to converge, it was necessary to include on the order of $10^6$ clusters of three and four nuclear spins in our Hahn-echo calculations for \ch{^{13}C} nuclear spins, and additionally $10^6$ clusters of five for \ch{^{14}N}. In contrast to the results obtained for the free nuclear spin bath ($m_s = 0$), the Hahn echoes for the NV center in the $m_s = -1$ state are identical with both the CCE and gCCE methods (See SI), indicating a complete suppression of the spin relaxation process.

Unlike the Hahn echo, the Ramsey signal remains mostly unchanged when the electron is in the $m_s = -1$, compared to that of the free nuclear spins. The $T_2^*$ is limited by the interactions with the small number of nearest bath spins \cite{PhysRevB.85.115303}. We note that each experimental data falls well within the computed distribution (Fig. \ref{fig:nonly}b); however the computed $T_2^*$ is overestimated for specific nuclear spins, likely due to the exclusion radius around the experimental cluster used to set up our model (see Methods). Our results point at an amount of nuclear spins in the proximity of each of the registers which is larger than expected based on the number of experimentally identified positions \cite{Abobeih2019}. 

Overall we find excellent agreement between experimental and computed values, thus validating the applicability and accuracy of our computational framework, even in the presence of the dominating hyperfine field of the electron spin.

\subsection{Coherence time of nuclear spins in the vicinity of an NV center}
Having validated our computational framework, we now turn to investigating the dependence of the nuclear coherence time on the position and orientation of the nuclear spins inside the frozen core (Fig. \ref{fig:t2map}a) of the NV center. 

In Figure \ref{fig:t2map}b, we report a complete map of the nuclear spins ensemble averaged $T_2$, computed as a function of the polar angle $\Theta$ and the distance from the electron spin $d$, for the NV center in the $m_s = -1$ state. We find that $T_2$ in the limit of high magnetic field ranges from more than 600 ms for nuclear spins within 0.5 nm of the electron spin to less than 50 ms for nuclear spins at distances larger than 4 nm. Figure \ref{fig:t2map}c reports cuts along several polar angles, showing a strong dependence of the coherence time on the orientation of the nuclear spin with respect to the NV center. The coherence time reaches its maximum for equatorial nuclear spins in the (111) plane, $\Theta=90^{\circ}$. In the vicinity of the NV, the distribution of computed $T_2$ values matches that of the spin density of the defect (Fig. \ref{fig:t2map}a), and the nuclear spins located where the spin density is the highest exhibit the maximum coherence time. In contrast, the axial nuclear spins along the [111] axis have longer $T_2$ times at larger distances.
Interestingly, our calculations show that at about 4 to 5 nm from the NV, $T_2$ is the same as in the absence of the electronic spin. This distance is smaller than the one ($d \ge 6$ nm) at which the average strength of the inter-nuclear interactions ($\sim 60$ kHz) exceeds that of the mean hyperfine coupling.

\begin{figure}
    \centering
    \includegraphics[scale=1]{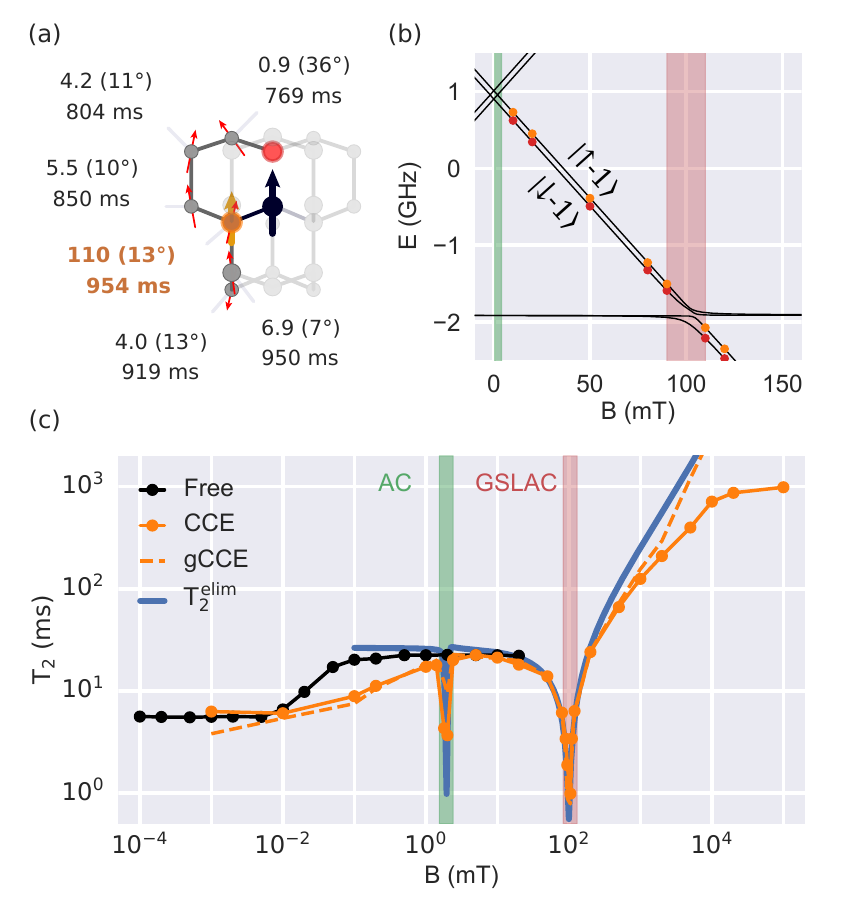}
    \caption{ (a) Computed first and second shell ensemble-averaged nuclear spin coherence times. For each nuclear spin, we also report the hyperfine coupling $\Tilde{A}_{zz} = \sqrt{A_{xz}^2 + A_{zz}^2}$ in MHz, the angle $\Theta_A$ between the [111] axis of diamond and the hyperfine quantization axis, and $T_2$ computed in the limit of a large magnetic field. For the first shell nuclear spin (highlighted in orange), the coherence time is computed at 10 T, and for all others at 1T. Red arrows show hyperfine quantization axes. (b) Energy levels of the hybrid electron-nuclear spins system for NV and first shell nuclear spin as a function of an applied magnetic field along the [111] direction. Orange and red dots correspond to $\ket{0_a}$ and $\ket{1_a}$ levels respectively (see text).  (c) Coherence time of the first-shell \ch{^{13}C} as a function of the applied magnetic field along the [111] axis, computed with CCE (solid orange line), gCCE (dashed orange line), and hybridization-limited $T_2^{\text{elim}}$ (blue line, see text). $T_2$ for the free nuclear spin (black) is shown as a comparison. The green shaded region denotes avoided crossing (AC) in the electronic levels due to the hyperfine interactions; the red shaded region denotes ground state level anticrossing (GSLAC) of the electronic levels.
}
    \label{fig:strong}
\end{figure}

We focus next on nuclear spins strongly coupled to the spin defect, with the interaction values in the MHz range. As an example, we consider \ch{^{13}C} (Fig. \ref{fig:strong}a) belonging to the first shell. We write the central spin Hamiltonian as:

\begin{equation}
\begin{split}\label{eq:conv_cce_ham}
    \hat H_{en} = D \hat{S_z}^2 + \gamma_{e} B_z \hat S_{z} + \gamma_{n} B_z \hat I_{z} + A_{zz}\hat S_{z}\hat I_z + \\ A_{xx}\hat S_x\hat I_x + A_{yy}\hat S_y\hat I_y + A_{xz}(\hat S_x\hat I_z + \hat S_z\hat I_x)
\end{split}
\end{equation}
Where $\hat S_i$, $\hat I_i$ are electron and nuclear spin operators, $D = 2.88$ GHz is the NV center zero-field splitting, $\gamma_{e}$, $\gamma_{n}$ are electron and \ch{^{13}C} nuclear spin gyromagnetic ratios respectively (see Methods for details). $A_{xx}=99.8$ MHz, $A_{yy}=176.8$ MHz, $A_{zz}=108.0$ MHz, $A_{xz}=25.5$ MHz are hyperfine couplings obtained from DFT calculations and are in good agreement with experimental data \cite{PhysRevB.94.060101}. The energy levels of the combined electron-nuclear spins system are shown in Figure \ref{fig:strong}b.


We find that to obtain saturation of the coherence time of the first shell \ch{^{13}C}, a much higher magnetic field is required than in the case of free nuclear spins (Fig. \ref{fig:strong}c). Strikingly, at low applied fields, the nuclear spin $T_2$  is severely affected by the partial hybridization of the electronic spin levels induced by the hyperfine coupling. To analyze this effect, we express the two energy levels $\ket{0_a}$ and $\ket{1_a}$ of the hybrid electron-spin nuclear spin system as:

\begin{equation}
    \ket{1_a} = \ket{-1\uparrow} + 
                \alpha^{1a}_{-1\downarrow} \ket{-1\downarrow} + 
                \alpha^{1a}_{0\uparrow} \ket{0\uparrow} + 
                \alpha^{1a}_{0\downarrow} \ket{0\downarrow} 
\end{equation}
\begin{equation}
    \ket{0_a} = \ket{-1\downarrow} + 
                \alpha^{0a}_{-1\uparrow} \ket{-1\uparrow} +  
                \alpha^{0a}_{0\uparrow} \ket{0\uparrow} + 
                \alpha^{0a}_{0\downarrow} \ket{0\downarrow} 
\end{equation}


In the limit of $B\rightarrow \infty$, the amplitudes $\alpha_i$ vanish. The latter can be computed by directly diagonalizing the Hamiltonian or from perturbation theory (see SI) and are expected to be significantly smaller than one. If the reduced density matrices of the electron spin in states $\ket{0_a}$ and $\ket{1_a}$ differ substantially, we expect a significant impact of the mixing of electron spin levels on the nuclear spin coherence time. To estimate the effect on T$_2$ of the difference in hybridization between the $\ket{0_a}$ and $\ket{1_a}$ levels, we use a modified approximate model first suggested in Ref. \cite{Balian2014}. The model was proposed to predict the $T_2$ of two electron-spin states with similar magnetization in the high-field regime, when slow oscillations of nuclear spin pairs dominate the decoherence process.
Using such a model, the contribution to the \textit{nuclear spin} coherence time arising only from the electronic hybridization (which we denote as electron-limited, (\text{elim})) can be expressed as:

\begin{equation}
 T_2^{\text{elim}}(B)\approx \mathcal{C} \frac{||P_{0a}(B)||+||P_{1a}(B)||}{||P_{0a}(B)-P_{1a}(B)||},
\end{equation}
where $P_{0a}(B)=\bra{0_a}\mathbf{S}\ket{0_a}$, $P_{1a}(B)=\bra{1_a}\mathbf{S}\ket{1_a}$ are the effective magnetization of the electron spin in the $\ket{0_a}$ and $\ket{1_a}$ states respectively, $\mathcal{C}$ is a magnetic field-independent constant, specific to a given system. We find $\mathcal{C}$ to be equal to 0.31 ms for the parameter range appropriate for the system under study (See SI). 
The electron-limited coherence obtained from the model agrees well with the predictions of the full quantum mechanical treatment over a wide range of magnetic fields, thus confirming the significant impact of the hybridization of the electron spin levels on the coherence time of nuclear spins.

Using perturbation theory, we obtain an approximate expression for the electron spin-limited coherence time (see SI):

\begin{equation}\label{eq:t2elim_approx}
 T_2^{\text{elim}}(B)\approx \frac{4 \mathcal{C} (D+\gamma_e B)(A_{zz}+\gamma_n B)}
 {A_{xz}(A_{xx}+2A_{zz}+2\gamma_n B)}
\end{equation}

We find that $T_2^{\text{elim}}$ is proportional to $\tan^{-1} (\Theta_A)$ at intermediate magnetic fields, where $\Theta_A$ is the angle between the hyperfine quantization axis $\mathbf{n}=A_{xz}\mathbf{i} +  A_{yz}\mathbf{j} + A_{zz}\mathbf{k}$ and the [111] direction of the diamond lattice. Hence our results show that the $T_2$ of any nuclear spin with a substantial perpendicular component of the hyperfine coupling requires a significantly higher magnetic field to achieve saturation when the NV is in $m_s=-1$ state. One can use Eq. \ref{eq:t2elim_approx} to estimate the conditions at which the impact of the hybridization of the electron spin levels on nuclear spin $T_2$ becomes insignificant.

The effect of the partial hybridization of electron spin levels has the highest impact near avoided crossings of energy levels. In the case of electron spins, avoided crossings originating from hyperfine interactions lead to a decoherence-protected subspace \cite{RamaKoteswaraRao2020}. Instead, the effect of these transitions on the $T_2$ of nuclear spins is extremely detrimental. At each of the avoided crossings, our calculations show a sharp dip in the coherence time of strongly coupled nuclear spins, highlighting the important trade-off one faces in utilizing nuclear spins as memory qubits at avoided crossings \cite{Ruskuc2022}.


\begin{figure}
    \centering
    \includegraphics[scale=1]{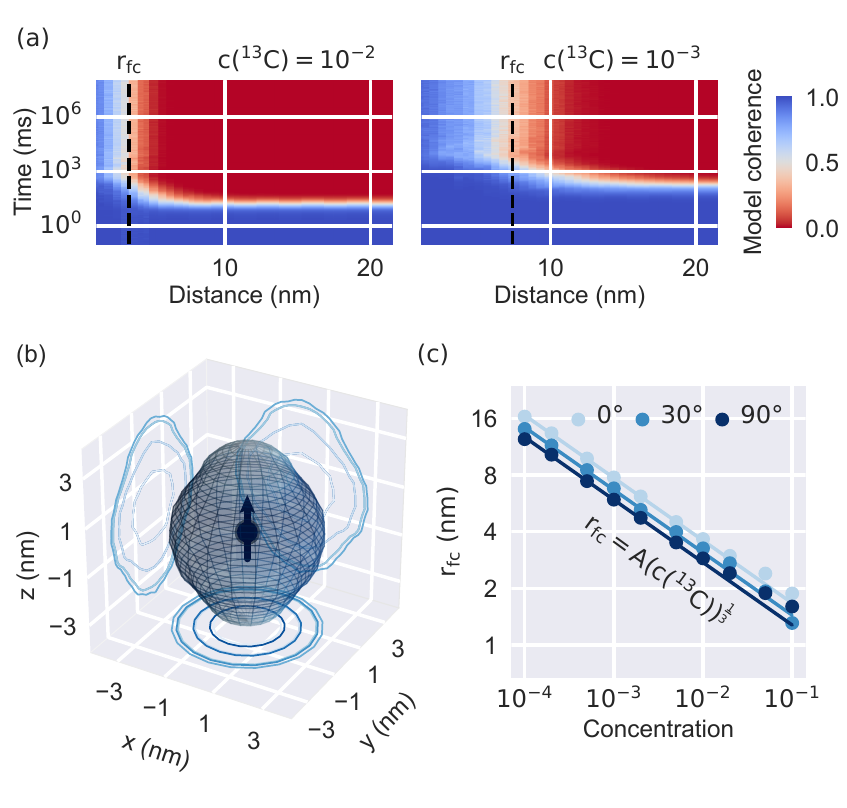}
    \caption{(a) Spin-pair coherence function as a function of distance from the NV for two different isotopic concentrations. The frozen core radius $r_{fc}$ is computed as the distance at which the model coherence function at time $\ge 10^6$ ms decreases to $\sfrac{1}{e}$ (b) Shape of the NV frozen core at natural isotopic concentration. (c) Scaling of the $r_{fc}$ with \ch{^{13}C} concentrations at different polar angles $\Theta$ (See Fig. \ref{fig:t2map}). Solid lines show fits to the function $r_{fc}=Ac^{-\sfrac{1}{3}}$.}
    \label{fig:frozencore}
\end{figure}


We close this section by presenting a model of the frozen core aimed at understanding its spatial extent. We use a simplified spin pair-only model, where the size of the bath and spin-pair cutoff radii are obtained in the absence of the electron spin (see Methods). We find that within this approximation, the computed Hahn echo of the nuclear spins in the vicinity of the NV center persists indefinitely. In the opposite limit of large distances between nuclear and electron spins, the model yields a coherence function decaying to zero, as expected (Fig. \ref{fig:frozencore}a). The distance from the NV at which the model coherence function changes its behavior from constant to decaying determines the boundary of the frozen core. Specifically, we define the frozen core radius ($r_{fc}$) as the distance at which the model coherence at an infinite time decreases to $\sfrac{1}{e}$.

Using this definition, we find that the frozen core of the NV center is asymmetrical and elongated along the z-axis (Fig. \ref{fig:frozencore}b). The radius $r_{fc}$ varies from 2.7 nm at $\Theta = 55 ^{\circ}$ to 3.8 nm at $\Theta = 180 ^{\circ}$, matching the coherence time behavior shown in Fig \ref{fig:t2map}c: the computed $T_2$ decays below 100 ms at 3.8 nm for axial and at 2.9 nm for equatorial spins nuclear spins. In contrast, the $r_{fc}$ dependence on the azimuthal angle is negligible. The total volume of the frozen core is 165 nm$^3$, which corresponds to about 300 \ch{^{13}C} nuclear spins on average.

The frozen core size is correlated with the strength of the parallel component of the hyperfine interaction with the electron spin. Within the point-dipole approximation, this interaction can be written as \cite{slichter_2011}:

\begin{equation}
     A_{zz} = - \frac{\mathcal{G}}{r^3}(3\cos^2{\Theta} - 1)
\end{equation}

Where $\mathcal{G}=\frac{\mu_0 \gamma_e \gamma_n \hbar }{4\pi}=7.60$ Hz nm$^{3}$ for \ch{^{13}C} nuclear spins. 
We note that for different systems (such as quantum dots in Si \cite{PhysRevB.91.214303}), other terms might dominate the hyperfine interactions, and one can expect different shapes of the frozen core. Interestingly, at the angle $\arccos{(\sfrac{1}{\sqrt{3}})}$ where the dipolar coupling vanishes, the value of $r_{fc} = 2.7$ nm is only slightly smaller than 2.8 nm, obtained for $\Theta = 90 ^{\circ}$.

We find that the isotopic purification of the system leads to an increased size of the frozen core, where $r_{fc}$ scales as the cubic root of the isotopic concentration in a wide range of spin densities (Fig. \ref{fig:frozencore}c). The ratio between $r_{fc}$ at different polar angles also remains constant. The scaling deviates from cubic only at very high concentrations, at which the actual electron spin density distribution and discrete lattice site positions  should be taken into account.


\begin{figure}
    \centering
    \includegraphics[scale=1]{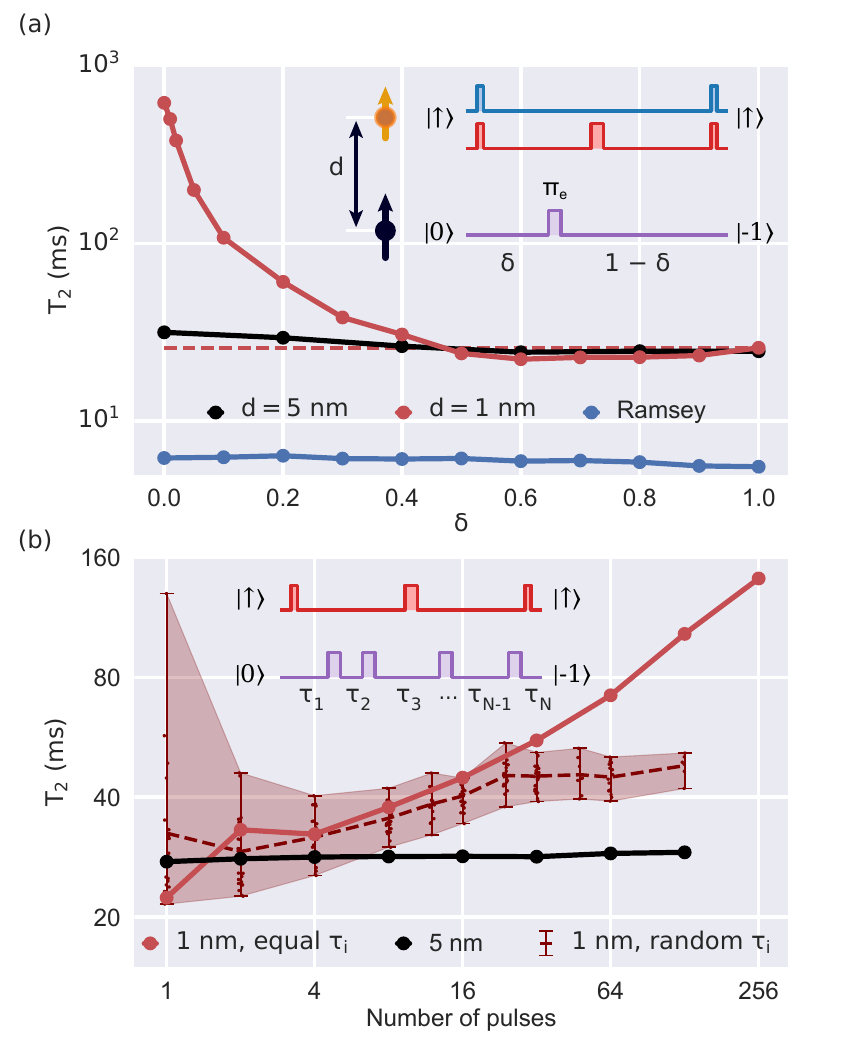}
    \caption{(a) Pulse sequences and the corresponding computed nuclear spin $T_2$ (black and red) and $T_2^*$ (blue) for a single nuclear spin at distances 1 nm and 5 nm from the NV center when a single $\pi_e$ pulse is applied to the electron spin. The dark blue (orange) arrow represents an electron (nuclear) spin. $T_2^*$ is shown for d = 1 nm.(b) Pulse sequence and the nuclear spin $T_2$ when many $\pi_e$ pulses are applied to the electron spin. Red color shows $T_2$ of the nuclear spin at 1 nm. Spacing between the pulses ($\tau_i$) is either random (points inside shaded area) or constant (solid line). Black line shows $T_2$ of the nuclear spin at 5 nm.}
    \label{fig:mpulses}
\end{figure}


\subsection{Effect of electron spin control on nuclear spin coherence}
Having analyzed the characteristics of nuclear spin coherence times in the vicinity of an NV center as a function of the central spin state, we now investigate how nuclear spin coherence is affected by {\it changes} in the state of the electronic spin. The dynamical change of the state of the NV has been shown to be a valuable tool for improving nuclear spin coherence. For example, one can use unbalanced echo \cite{unbalecho} by applying control pulses to the electron spin to enhance the protection of the nuclear spin ensembles against lattice strain noise.

We analyze the effect of the central spin coherent control on nuclear spin coherence by applying a sequence of $\pi_e$-pulses (here, the index $e$ denotes electron spin) to the NV center and we compute the dynamics of nuclear spins as a function of distance from the NV.
 
Figure \ref{fig:mpulses}a shows the nuclear spin coherence when a single $\pi_e$-pulse is applied to the electron spin at different fractions of the total time $0 \le \delta \le 1$. The electron spin is initialized in the $m_s=0$ state; after the $\pi_e$-pulse is applied, the electron spin rotates into the $m_s=-1$ state.
Before the $\pi_e$ pulse, the nuclear spin precesses with  frequency $w_L^{(0)}=-\gamma_n B$; upon the application of the pulse, the frequency is   $w_L^{(-1)}=-\gamma_n B - A_{zz}$, leading to the emergence of a nonzero phase of the Hahn-echo signal. In our calculations, we obtain the decay of the coherence time from the absolute value of the coherence function $|\mathcal{L}|$.


We find that the coherence time of the nuclear spins outside the frozen core does not change significantly with the state of the central spin. However, within the frozen core, the change is drastic: when $\delta > 0.5$, the nuclear spin coherence time reaches a local minimum and we observe a 15\% drop in $T_2$, compared to that of the spin in the $m_s = 0$ state, indicating a destructive interference between nuclear and electron control pulses.

Figure \ref{fig:mpulses}b shows the $T_2$ of the nuclear spin as a function of the number of applied $\pi_e$ pulses. The electron spin is initialized in the $m_s = 0$ state. In this case, one can achieve a so-called motional narrowing of the hyperfine field \cite{Ruskuc2022}: as the number of $\pi_e$ pulses increases, the electron-induced field rapidly oscillates and its overall effect can be described by an average field. The motional narrowing leads to a significant enhancement in coherence time. We obtain the highest increase in T$_2$ for a constant spacing between $\pi_e$ pulses; however, $T_2$ is still much smaller than the coherence time achieved when the electron spin remains in the $m_s = -1$ state (620 ms).


\section{Discussion and Outlook}
In this work, we presented and validated a robust computational protocol to estimate nuclear spin coherence times. In the absence of electron spins, we find nuclear spin coherence times to be limited almost equally by dephasing and relaxation processes. However, the interplay between these two processes greatly varies depending on the specific spatial configuration of nuclear spins. This finding indicates that the geometrical arrangements of the nuclear spin environment of spin defects should be carefully characterized \cite{Abobeih2019}, in order to identify the optimal nuclear spins to store nuclear quantum states as long as possible.
Our calculations showed that the Hahn echo of nuclear spins exhibit complex oscillatory features emerging from the spin exchange interactions with the bath. These oscillations arise from the direct interactions between a single bath spin and the central spin, and they can be used to identify and characterize spin exchange interactions with the single spins in the bath.

In the presence of  electron spins, we characterized the shape of the frozen core of nuclear spins around the defect with a spin-pair model. The core turns out to be oblong and elongated along the $z$-axis, matching the dependence of the hyperfine coupling on the polar angle. We find that the volume of the frozen core is inversely proportional to the concentration of nuclear spins; thus, the total number of nuclear spins inside the frozen core is constant and equal to about 300 \ch{^{13}C}, irrespective of any isotopic purification.
We analyzed the frozen core effect on coherence times and found that the Hahn-echo coherence time of the nuclear spins can be enhanced by up to 36 times for the closest nuclear spins, when an electron is in the $m_s=-1$ state. Near an NV center, the highest $T_2$ is attained by equatorial nuclear spins, closely matching the spin density distribution of the NV center. Further away from the electronic spin defect, it is the polar nuclear spins that retain the highest $T_2$. Our calculations showed the striking effect of electronic orbital hybridization on nuclear coherence times and the negative impact of avoided crossings. We also showed that strongly coupled nuclear spins require significantly higher magnetic fields to suppress the effect of partial hybridization of electron spin levels on their coherence time.

Finally, we uncovered that the coherent NV center control, leading to a change of electron magnetic states, severely impacts the nuclear spin coherence time. Even with no noise affecting the NV state, we find that the nuclear $T_2$ inside the frozen core is dramatically decreased as soon as the state of the NV center is changed.

Overall, the validated computational framework proposed here for the study of nuclear spin registers is general and applicable to broad classes of systems and problems. For example, one can use the proposed platform to study the impact of the total spin of an electron qubit on the nuclear spins' frozen core. In particular, one could investigate the difference in coherence times in the presence of electron spin-$\sfrac{1}{2}$ qubits, exhibiting a hyperfine field in any state, and NV centers, where different magnetic states have different hyperfine coupling. Importantly, using our computational platform one may screen materials for optimal nuclear spin coherence times\cite{Kanai2022}.

Another interesting avenue of research is the exploration of the predicted frozen core size and shape as an engineering tool for the bottom-up design of spin qubits in molecular systems \cite{Laorenza2022}. With each frozen core corresponding to a computational domain of a specific electron spin, one can envision a nanoscale network of spin processors, with electron spins as processing units and nuclear spins acting as memory qubits.

Finally, our results tell a series of cautionary tales for the applications of nuclear spins for quantum technologies. From the applied magnetic field to the NV control, we elucidated the various noise channels that may adversely affect the quantum state of the nuclear spins.

\section{Methods}

The quantum evolution of the combined electron spin-nuclear register is described by the model Hamiltonian:

\begin{equation} \label{Total H}
\hat H = \hat H_{en} + \hat H_{en\text{-}b} + H_{b}.
\end{equation}

The central spin Hamiltonian $\hat H_{en}$ includes:
\begin{equation} \label{en H}
\hat H_{en} = D\hat S_z + \gamma_{e} \textbf{B}\cdot\textbf{S} + \gamma_{n} \textbf{B}\cdot\textbf{I}_0 + \textbf{S}\cdot\textbf{A}_0\cdot\textbf{I}_0.
\end{equation}

Here $D$ is the zero field splitting of the electron spin, $\textbf{B}=(B_x, B_y, B_z)$ is the magnetic field, $\gamma_{n}$ is the gyromagnetic ratio of the \ch{^{13}C} nuclear spin, $\textbf{S} = (\hat S_x, \hat S_y, \hat S_z)$ and $\textbf{I}_i = (\hat I_{x, i}, \hat I_{y, i}, \hat I_{z, i})$ denote electron and the $i$-th nuclear spin operators, respectively. The zero index denotes a given  nuclear spin chosen as a qubit. 

The bath-central spins Hamiltonian $\hat H_{en\text{-}b}$ and the bath Hamiltonian  $\hat H_{b}$ are defined as follows:

\begin{equation} \label{enB H}
   \hat H_{en\text{-}b}= \sum_i{\textbf{S}\cdot\textbf{A}_i\cdot\textbf{I}_i +  \textbf{I}_0\cdot\textbf{P}_{0i}\cdot\textbf{I}_i},
\end{equation}
and
\begin{equation} \label{B H}
   \hat H_{b} = \sum_i{-\gamma_{n} \textbf{B}\cdot\textbf{I}_i} + \sum_{i \ge j} \textbf{I}_i\cdot \textbf{P}_{ij} \cdot\textbf{I}_j.
\end{equation}

Here $\textbf{A}_i$ is the hyperfine coupling tensor of the $i$-th nuclear spin, and $\textbf{P}_{ij}$ is the dipole-dipole coupling between spins $i$ and $j$.

We use the cluster-correlation expansion (CCE) to compute the coherence function of the nuclear spin, defined as:
\begin{equation}
\mathcal{L}(t) = \frac{\langle \hat I_-(t)\rangle}{\langle \hat I_-(0)\rangle} = \frac{\bra{\uparrow}\hat \rho (t) \ket{\downarrow}}{\bra{\uparrow}\hat \rho (0) \ket{\downarrow}},
\end{equation}
where $\ket{\uparrow}$ and $\ket{\downarrow}$ are nuclear spin-up and spin-down states and $\hat I_-$ are nuclear lowering spin operators, $\hat \rho (t)$ is the density matrix of the central spin. In the presence of an NV center we define off-diagonal elements between eigenstates of $\hat H_{en}$ corresponding to the diabatic levels $\ket{\uparrow 0}$, $\ket{\downarrow -1}$ for $m_s=0$, and to $\ket{\uparrow -1}$, $\ket{\downarrow -1}$  for $m_s=-1$ cases.

The CCE method up to the second order was previously used to qualitatively investigate the properties of the nuclear spins in procimity of shallow donors in Si \cite{PhysRevB.91.214303}. Here we apply the fully converged CCE and generalized CCE schemes with "externally aware" cluster corrections \cite{PhysRevB.86.035452} to quantitatively reproduce  the experimental data. 

Within the CCE scheme, the coherence function $\mathcal{L}(t) $ is factorized into the contributions of bath spin clusters with different size \cite{PhysRevB.78.085315}:

\begin{equation}\label{gCCE_eq}
\mathcal{L}(t) = 
            \prod_{i} \tilde L^{\{i\}}
            \prod_{i, j} \tilde 
            L^{\{ij\}} ...
\end{equation}

The contributions are computed recursively from the coherence function of the central spin, interacting with only a given cluster $C$ as $\Tilde{L}_C = \frac{L_{C}}{\prod_{C'}\Tilde{L}_{C'\subset C}}$, where the subscript $C'$ indicates all sub-clusters of $C$.

Depending on the framework, the $L_{C}$ are computed as follows. In conventional CCE \cite{PhysRevB.78.085315, PhysRevB.79.115320} (referred throughout the text as CCE), the relaxation processes of the central spin are discarded, and the coherence function is computed as an overlap in the cluster evolution, dependent on the central spins state:
\begin{equation}\label{eq:l_contribution}
    L_{C} = \bra{C}\hat U_C^{(0)}(t) \hat U_C^{(1) \dagger}(t)\ket{C},
\end{equation}
where $\ket{C}$ is the initial state of the cluster $C$. $\hat U_C^{(\alpha)}(t)$ is the time propagator defined in terms of the effective Hamiltonian $\hat H_C^{(\alpha)}$ conditioned on the qubit levels. Up to the second order of perturbation theory it can be written as:
\begin{equation}
\hat H_C^{(\alpha)} = \bra{\alpha}\hat H_C \ket{\alpha} + \sum_{i\neq\alpha}\frac{\bra{\alpha}\hat H_{b}\ket{i}\bra{i}\hat H_{b}\ket{\alpha}}{E_\alpha-E_i},
\end{equation}
where $\ket{\alpha}$, $\ket{i}$ are eigenstates of the central spins Hamiltonian $\hat H_{en}$, $\hat H_C$ is the Hamiltonian in Eq. (\ref{en H}) including only the bath spins in the cluster $C$:

\begin{equation} \label{Cluster H}
\hat H_C = \hat H_{en} +  \hat H_{en\text{-}b}^{(i\in C)} +  \hat H_{b}^{(i,j\in C)}
\end{equation}

In contrast, in the generalized CCE (gCCE) we compute the cluster contributions from the respective elements of the reduced density matrix of the central spin as \cite{Onizhuk2021c}:
\begin{equation}
    L_{C} = \bra{a} \Tr_C[\hat \rho_{en\otimes C} (t)]\ket{b},
\end{equation}
where $\rho_{en\otimes C} (t)$ is the density matrix of the system, which includes bath spins in the cluster $C$ and all central spins. The evolution is computed using the full cluster Hamiltonian $\hat H_C$. 

We use Monte Carlo bath state sampling \cite{Onizhuk2021c} which has been shown to improve the convergence of CCE method in the all-dipolar spin systems \cite{PhysRevB.86.035452}.

We use the PyCCE module \cite{Onizhuk2021b} to carry out all CCE simulations. To approximate the dipolar coupling parameters,
we use the actual spin density of the NV center in diamond, computed with density functional theory at the PBE \cite{PhysRevLett.77.3865} level in a 1000 atoms supercell using the Quantum Espresso package \cite{Giannozzi2009}. The dipolar coupling is then computed as \cite{Ghosh2019}:
\begin{equation}
    A_{ab}=\frac{1}{2}\frac{\mu_0}{4\pi}\gamma_e\gamma_n\hbar^2\int{\frac{|\textbf{r}|^2\delta_{ab}-3\textbf{r}_a\textbf{r}_b}{|\textbf{r}|^5}\rho_s{(\textbf{r})}d\textbf{r}},
\end{equation}
where $\textbf{r}$ is the position relative to a  given nuclear spin, $\rho_s$ is the electron spin density.

The contact terms of the nuclear spins at distances under 1 nm were computed using the GIPAW module of Quantum Espresso. For every other nuclear spin we assumed the contact terms to be vanishing.

To approximate the experimental nuclear spin bath we generated 50 random bath configurations around the experimental cluster. 
Assuming that all closest nuclear spins were identified in the experiment, we imposed a cutoff of $0.56$ nm around each of the identified nuclear spins. This cutoff is chosen so that the exclusion volume on average contains 27 nuclear spins.

To characterize the extent of the frozen core, we used the CCE2 (spin-pair approximation) with the number of pairs converged in the absence of the electron spin. Using the same number of pairs, we computed the hypothetical coherence signal at various distances from the NV center.

In the simulations all $\pi$-pulses are assumed to be ideal, instantaneous, and selective to the spin chosen as a central one.

In the experiment, to make the $\pi$-pulses selective to a specific nuclear spin, one has to rotate NV into $m_s=-1$ state for a short period of time, regardless of the NV state of interest. This means that even the data for $m_s=0$ state includes a small fraction of time during which the NV is in $m_s=-1$ state. Depending on the nuclear spin, this time is between 0.3-1.6 ms (See ref.\cite{10qubitreg} Table S3 for specific gate times), which can lead to the nuclear spin acquiring an additional phase, not present in the simulations.




%

\section{Acknowledgements}
This work was supported by the Design and Optimization of Synthesizable Materials with Targeted Quantum Characteristics (AFOSRFA9550-19-1-0358). M.O. acknowledges the support of a Google PhD Fellowship. We thank C. E. Bradley and T. H. Taminiau for providing the experimental data, and C.P. Anderson, M. Raha, Y. Wang for useful comments. 

\section{Author contributions}
G.G. and M.O. conceived the study. M.O. developed the theoretical framework, performed simulations, and analysed the data. Both authors contributed to the writing of the manuscript.

\section{Competing Interests}
The Authors declare no Competing Financial or Non-Financial Interests.


\end{document}